\documentclass[aps,twocolumn,showpacs,superscriptaddress]{revtex4}

\usepackage{epsfig}
\usepackage{graphicx}
\usepackage{amsfonts}
\usepackage[figuresright]{rotating}
\usepackage{amssymb}
\usepackage{amsmath}
\usepackage{psfrag}

\def\be{\begin{equation}}
\def\ee{\end{equation}}
\def\bea{\begin{eqnarray}}
\def\eea{\end{eqnarray}}

\def\nn{\nonumber}

\begin{document}
\title{The in-plane gradient magnetic field induced vortex lattices in
spin-orbit coupled Bose-Einstein condensations}

\author{Xiang-Fa Zhou}
\affiliation{Key Laboratory of Quantum Information, University of
Science and Technology of China, Chinese Academy of Sciences, Hefei,
Anhui 230026, China}
\affiliation{Synergetic Innovation Center of Quantum Information
and Quantum Physics, University of Science and Technology of China, Hefei,
Anhui 230026, China}

\author{Zheng-Wei Zhou}
\affiliation{Key Laboratory of Quantum Information, University of
Science and Technology of China, Chinese Academy of Sciences,
Hefei, Anhui 230026, China}
\affiliation{Synergetic Innovation Center of Quantum Information
and Quantum Physics, University of Science and Technology of China, Hefei,
Anhui 230026, China}

\author{Congjun Wu}
\affiliation{Department of Physics, University of California, San Diego,
CA 92093, USA}

\author{Guang-Can Guo}
\affiliation{Key Laboratory of Quantum Information, University of Science
and Technology of China, Chinese Academy of Sciences, Hefei,
Anhui 230026, China}
\affiliation{Synergetic Innovation Center of Quantum Information and
Quantum Physics, University of Science and Technology of China, Hefei,
Anhui 230026, China}

\begin{abstract}
We consider the ground-state properties of the two-component spin-orbit
coupled ultracold bosons subject to a rotationally symmetric in-plane gradient magnetic field.
In the non-interacting case, the ground state supports 
giant-vortices carrying large angular momenta without rotating the trap.
The vorticity is highly tunable by varying the amplitudes and
orientations of the magnetic field.
Interactions drive the system from a giant-vortex state to various
configurations of vortex lattice states along a ring.
Vortices exhibit ellipse-shaped envelops with the major and minor axes
determined by the spin-orbit coupling and healing lengths, respectively.
Phase diagrams of vortex lattice configurations are constructed and
their stabilities are analyzed. 
\end{abstract}
\maketitle


\section{introduction}

Spin-orbit (SO) coupling plays an important role in contemporary
condensed matter physics, which is linked with many important effects
ranging from atomic structures, spintronics, to topological insulators
\cite{zutic2004,Hasan2010,Qi2011}.
It also provides a new opportunity to search for novel states with
ultracold atom gases which cannot be easily realized in
condensed matter systems.
In usual bosonic systems, the ground state condensate wavefunctions
are positive-definite known as the ``no-node'' theorem \cite{feynman1972,
wu2009}.
However, the appearance of SO coupling invalidates this theorem \cite{wu2011}.
The ground state configurations of SO coupled Bose-Einstein condensations (BEC)
have been extensively investigated and a rich structure exotic phases
are obtained including the ferromagnetic and spin spiral condensations
\cite{wu2011,stanescu2008,wang2010,ho2011}, spin textures of the
skyrmion type \cite{wu2011,hu2012,sinha2011,li2012,kawakami2012},
and quantum quasi-crystals \cite{gopalakrishnan2013},  etc.
On the experiment side, since the pioneering work in the NIST
group \cite{lin2009}, it has received a great deal of
attention, and various further progresses
have been achieved \cite{zhang2012v1,wang2012,cheuk2012,qu2013,olson2014}.
Searching for novel quantum phases in this highly tunable system is
still an on-going work both theoretically and practically
\cite{wilson2013,achilleos2013,kartashov2013,ozawa2013,deng2012,
zhang2012,lobanov2014,luo2014}, which has been reviewed in
\cite{zhou2013, dalibard2011, zhai2014, galistki2013, goldman2013}.

On the other hand, effective gradient magnetic fields have been
studied in various neutral atomic systems recently.
For instance, it has been shown in Ref. \cite{anderson2013,xu2013}
that SO coupling can be simulated by applying a sequence of gradient
magnetic field pulses without involving complex atom-laser coupling.
In optical lattices, theoretic and experimental progresses show
that SO coupling and spin Hall physics can be implemented without
spin-flip process by employing gradient magnetic field
\cite{kennedy2013,aidelsburger2013}.
This represents the cornerstone of exploring rich many-body
physics using neutral ultracold atoms.
Additionally, introducing gradient magnetic fields has also been
employed to create various topological defects including
Dirac monopoles \cite{pietila2009} and knot solitons \cite{kawaguchi2008}.
It would be very attractive to investigate the exotic physics by
combining both SO coupling and the gradient magnetic field
together in ultracold quantum gases.

In this work, we consider the SO coupled BECs subject to an in-plane gradient magnetic field in a $2$D geometry.
Our calculation shows that this system support a variety of interesting phases.
The main features are summarized as follows.
First, the single-particle ground states exhibit giant vortex states carrying large angular momenta.
It is very different from the usual fast-rotating BEC system, in which the giant vortex state appears only as meta-stable states
\cite{schweikhard2004,mueller2002}.
Second, increasing the interaction strength causes the phase transition into the vortex lattice state along a ring plus a giant core.
The corresponding distribution in momentum space changes from a symmetric structure at small interaction strengths to an asymmetric one as the interaction becomes strong.
Finally, the size of a single vortex is determined by two different length scales, namely, the SO coupling strength together with the 
healing length.
Therefore, the vortex exhibits an ellipse-shaped envelope
with the principle axes determined by these two scales.
This is different from the usual vortex in rotating BECs
\cite{fetter2009,zhou2011,xu2011,radic2011,aftalion2013,fetter2014},
where an axial symmetric density profile is always favored.

The rest of this article is organized as follows.
In Sect. \ref{sect:ham}, the model Hamiltonian is introduced.
The single particle wavefunctions are described
in Sect. \ref{sect:single}.
The phase transitions among different vortex lattice configurations
are investigated in Sect. \ref{sect:interaction}.
The possible experimental realizations are
 discussed in Sect. \ref{sect:experiment}.
Conclusions are presented in Sect. \ref{sect:conclusions}.

\section{The model Hamiltonian}
\label{sect:ham}

We consider a quasi-$2$D SO coupled BEC subject to a spatially
dependent magnetic field with the following Hamiltonian as
\begin{eqnarray}
H=\int d \vec{r}^2 && \hspace{-.45cm} \hat{\psi}(\vec{r})^{\dag}
\Big\{  \frac{\vec{p}^2}{2m} + \Lambda  r \left ( \cos\theta \hat{r}
+\sin\theta \hat{\varphi} \right ) \cdot \vec{\sigma} \nn \\
&+& \frac{1}{2}m\omega^2r^2 \Big \}
\hat{\psi}(\vec{r})+H_{soc}+H_{int},
\label{eq:ham_1}
\end{eqnarray}
where $\hat{r}=\vec{r}/r$ with $\vec{r}=(x, y)$;
$\vec{\sigma}=(\sigma_x, \sigma_y)$ are the usual Pauli matrices;
$m$ is the atom mass;  $\omega$ is the trapping frequency;
$\Lambda$ is the strength of the magnetic field, and
$\theta$  denotes the relative angle between
the magnetic field and the radial direction $\hat r$.
Physically, this quasi-2D system can be implemented by imposing a
highly anisotropic harmonic trap potential
$V_H=\frac{1}{2}m(\omega^2 r^2+\omega^2_zz^2)$.
When $\omega_z \gg \omega$, atoms are mostly confined in the $xy$-plane,
and the wavefunction along $z$ axis is determined as a harmonic
ground state with the characteristic length $a_z=\sqrt{\hbar/(m\omega_z)}$.

For simplicity, the SO coupling employed below
has the following symmetric form as
\begin{eqnarray}
H_{soc}&=&\int d \vec{r}^2 \hat{\psi}(\vec{r})^{\dag}
\Big[ \frac{\lambda}{m} (p_x \sigma_x +p_y \sigma_y)
\Big]\hat{\psi}(\vec{r})  \nn
\end{eqnarray}
with $\lambda$ the SO coupling strength. 
We note that due to this term, the magnetic fields which couples to spin can be employed 
as a useful method to control the orbit degree of freedom of the cloud. 
The interaction energy is written as
\begin{eqnarray}
H_{int}=\frac{g_{2D}}{2} \int d \vec{r}^2 \hat{\psi}(\vec{r})^{\dag}\hat{\psi}(\vec{r})^{\dag}\hat{\psi}(\vec{r})\hat{\psi}(\vec{r}).
\end{eqnarray}
Here the contact interaction between atoms in bulk
is $g=4\pi\hbar^2 a_s/m$, where $a_s$ is the scattering length.
For the quasi-2D geometry that we focus on, the effective
interaction strength is modified as $g_{2D}=g_{3D}/(\sqrt{2\pi}a_z)$.

\section{single-particle properties}
\label{sect:single}

\begin{figure}
{\includegraphics[width=0.8\linewidth]{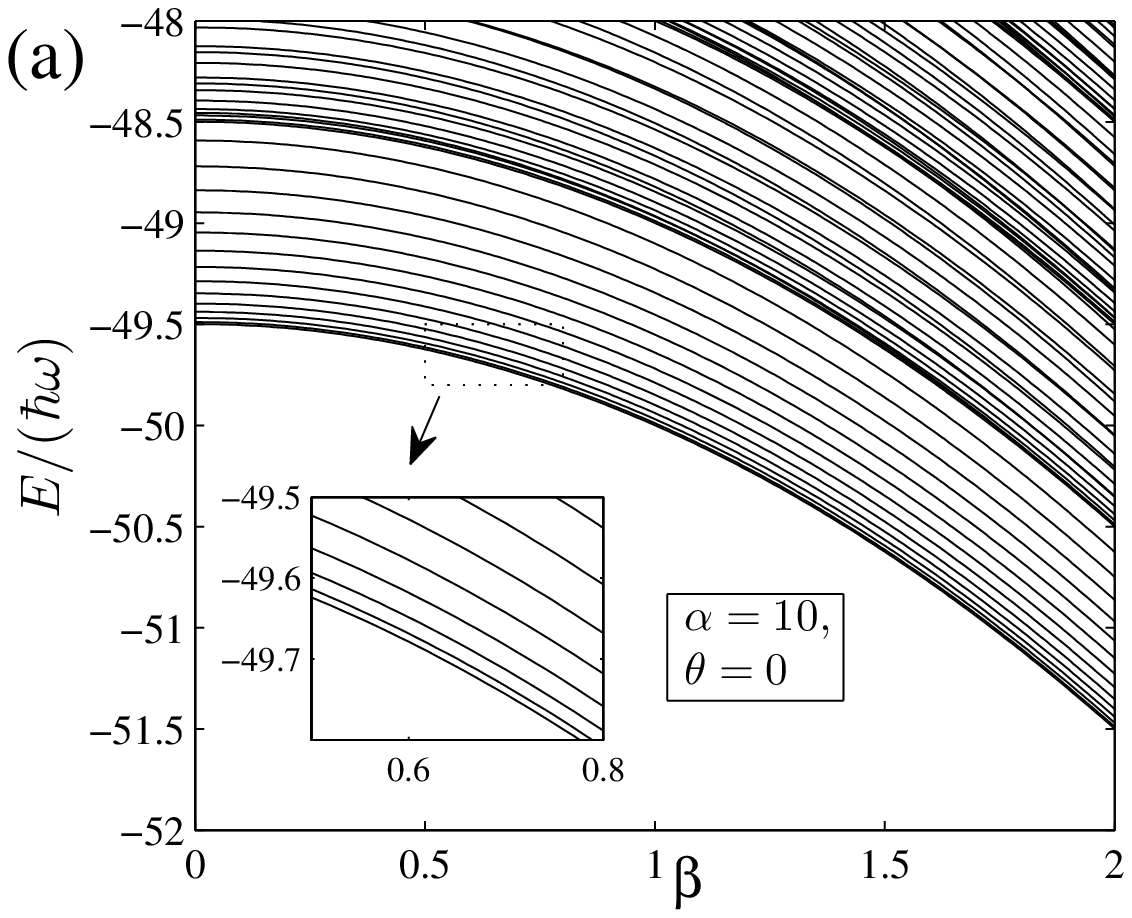}
\includegraphics[width=0.8\linewidth]{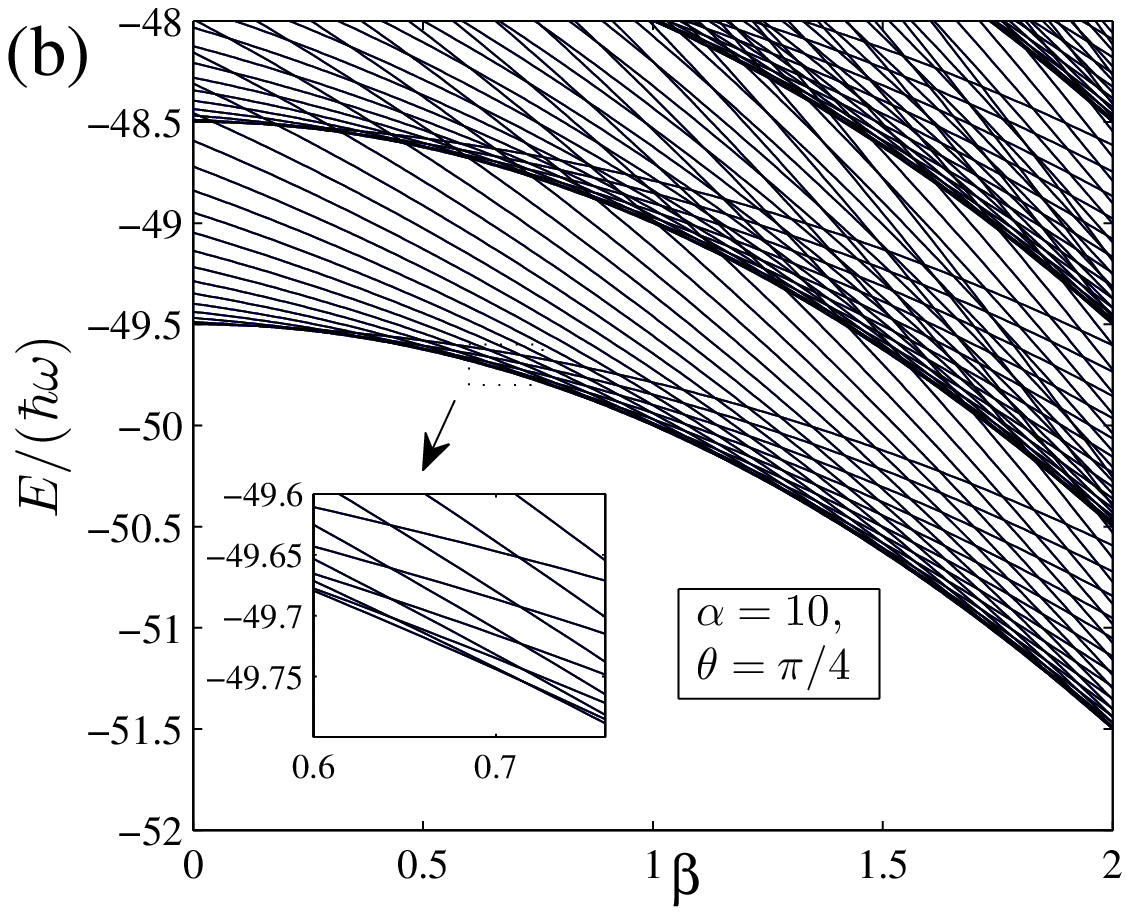}}
\caption{The single-particle dispersion of the Hamiltonian Eq.
(\ref{reducedsingleparticleH}) with lower energy branch as a
function of the reduced magnetic fields $\beta$ for fixed
$\alpha=10$ and different values of $\theta=0$ (a),
and $\frac{1}{4}\pi$ (b).
The inset in (b) shows that the ground states crossing for certain
values of $\beta$ at $\theta=\pi/4\neq0$, while there is no crossing
in (a) at $\theta=0$.
\label{spectrum}}
\end{figure}

The physics of Eq. \ref{eq:ham_1} can be illustrated by considering the
single-particle properties first.
After introducing the characteristic length scale of the confining
trap $l_T=\sqrt{\hbar/m\omega}$, the dimensionless Hamiltonian is
rewritten as
\begin{eqnarray}
    \frac{H_0}{\hbar\omega}=\int d \vec{\rho}^2 && \hspace{-.45cm} \hat{\phi}(\vec{\rho})^{\dag} \Big \{ -\frac{\vec{\nabla}^2}{2}  + \beta \rho \left ( \cos\theta \hat{r} +\sin\theta \hat{\varphi} \right ) \cdot \vec{\sigma} \nn \\
    &+&  \alpha \vec{k} \cdot \vec{\sigma} + \frac{1}{2}\rho^2 \Big \}\hat{\phi}(\vec{\rho}), \label{reducedsingleparticleH}
\end{eqnarray}
where $\alpha=\lambda/(m\omega l_T)$ and $\beta=\Lambda l_T/(\hbar \omega)$ are
the dimensionless SOC and magnetic field strengths, respectively;
the normalized condensates wave-function is defined as
$$\phi(\vec{\rho})=\frac{l_T}{\sqrt{N}}\Psi(\vec{r}=l_T\vec{\rho})$$
with $N$ the total number of atoms;

Since the total angular momentum $\hbar j_z=\hbar l_z+\frac{\hbar}{2}\sigma_z$
is conserved for this typical Hamiltonian, we can use it to label the
single-particle states.
If the magnetic field along the radial direction, i.e.,
$\theta=l \pi$,
the Hamiltonian also supports a generalized parity symmetry described by
$i\sigma_y P_x$, namely
\bea
[H_0, i\sigma_y P_x]=0,
\label{eq:symm}
\eea
with $P_x$ the reflection operation about the $y$-axis
satisfying $P_x: (x,y)\rightarrow (-x,y)$.
Therefore for given eigenstates $\phi_m=[f(\rho) e^{i m \varphi} ,
g(\rho) e^{i (m+1) \varphi}]^T$ with $j_z=(m+1/2)$, the above symmetry
indicates that these two states $\{\phi_m, (i\sigma_y P_x)\phi_m\}$
are degenerate for $H_0(\theta=l \pi)$.
This symmetry is broken when $\theta\neq l \pi$.

Due to the coupling between the real space magnetic field and momentum
space SO coupling, the single particle ground states exhibit interesting
properties at large values of $\alpha$ and $\beta$.
In momentum space, the low energy state moves to a circle with the
radius determined by $\alpha$.
The momentum space single-particle eigenstates break into two bands
$\psi^{\pm}(\vec k)$ with the corresponding eigenvalues
$E^{\pm}_{\vec{k}}/(\hbar \omega)=\frac{1}{2}
(|\vec{k}|^2\pm 2 \alpha |\vec{k}|)$
and eigenstates $\frac{1}{\sqrt 2}[1,\pm e^{i\theta_k}]^T$, respectively.
For the lower band which we focus on, the spin orientation
is $\langle\vec{\sigma}\rangle=(-\cos\theta_{\vec{k}}, -\sin\theta_{\vec{k}})$,
which is anti-parallel to $\vec{k}$.
On the other hand, in the real space, for a large value of $\beta$,
the potential energy in real space is minimized around the circle with
the the radius $r/l_T=\beta$ with a spatial dependent spin polarization.
Therefore around this space circle, the local wavevector at a position
$\vec r$ is aligned along the direction of the local magnetic field
to minimize the energy.
The projection of the local wavevector along the tangent direction
of the ring gives rise to the circulation, and thus the ground state
carries large angular momentum $m$ which is estimated as
\bea
m\simeq2\pi\beta\sin\theta/(2\pi/\alpha)=\alpha\beta\sin\theta.
\eea
Therefore, by varying the angle $\theta$, a series of ground
states are obtained with their angular momentum ranging from $0$ to
$\alpha\beta \gg 1$.
This is very different from the usual method to generate giant vortex,
where fast rotating the trap is needed \cite{fetter2009}.

\begin{figure}
{\includegraphics[width=\linewidth]{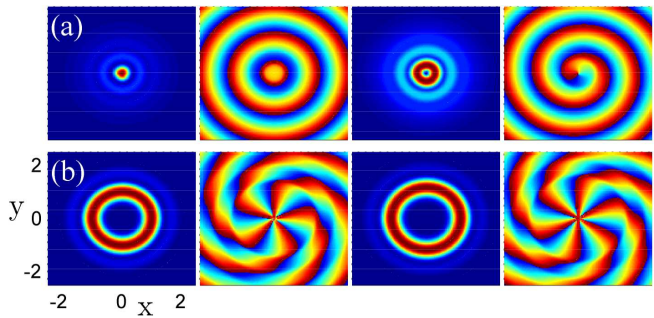}}
\caption{The density and phase profiles of the single-particle
ground states for fixed $\alpha=6$, $\beta=1$, and different
$\theta=\frac{1}{40} \pi$ (a), $\frac{2}{5}\pi$ (b).
From left to right: the density and phase profiles for
spin-up and spin-down components, respectively.
\label{densityphaseSP}
}
\end{figure}

For $\beta \gg 1$, the low energy wavefunctions mainly distribute around
the circle $\rho=\beta$.
As shown in Appendix \ref{sect:dispersion},
the approximated wavefunctions for the lowest band ($n=1$) is written as
\begin{eqnarray}
\phi_{n=1,j_z}(\rho,\varphi) &\simeq& \frac{1}{2\pi^{\frac{3}{4}}\rho^{\frac{1}{2}}}
e^{-\frac{(\rho-\beta)^2}{2}}e^{i\rho\alpha \cos\theta } \nonumber \\
&\times&
\left [ \begin{array}{c} e^{i [m\varphi-\frac{\theta}{2}]} \\
-e^{i [(m+1)\varphi+\frac{\theta}{2}]}  \end{array} \right ],
\end{eqnarray}
where $\varphi$ is the azimuthal angle.
The corresponding energy dispersion is approximated as
\begin{eqnarray}
E_{n,j_z}\approx n+\frac{1-\alpha^2-\beta^2}{2}
+ \frac{(j_z-\alpha\beta\sin\theta)^2}{2(\alpha^2\cos^2\theta+\beta^2)}.
\label{eq:dispersion}
\end{eqnarray}
For given values of $\alpha$ and $\beta$, $E_{n,j_z}$ is minimized
at $j_z \simeq \alpha\beta\sin\theta$, which is consistent with the
above discussion.
In the case of $\theta= l \pi$,  two states with $m=l$ and $-(l+1)$
are degenerated due to the symmetry defined in Eq. \ref{eq:symm}.
Interestingly, Eq. \ref{eq:dispersion} also indicates that for
integer $\alpha\beta\sin\theta=l$, an approximate degeneracy
occurs for $m=l$ and $l-1$.

Fig. \ref{spectrum} shows the single-particle dispersion of different
angular momentum
eigenstates along with the radius $\beta$ for different values of $\theta$.
For $\theta=0$, the dispersion with different $j_z$ never cross each other
Fig. \ref{spectrum} (1a).
The values of $j_z$ for the ground state are always $j_z=\frac{1}{2}$
or $-\frac{1}{2}$  due to the symmetry Eq. \ref{eq:symm}.
When $\theta=\pi/4\neq0$, the spectra cross at certain parameter
values, and the ground-state can be degenerate even without additional
symmetries as shown in Fig. \ref{spectrum} (1b), which is consistent
with above discussions.
For $\beta \gg 1$, the probability density of the ground state
single particle wavefunction mainly distributes around
a ring with $\rho=\beta$.
Interestingly, the phase distribution exhibits the typical
Archimedean spirals with the equal-phase line satisfying
$\rho \sim m \varphi$ (or $\rho\sim(m+1) \varphi$)
(see Fig. \ref{densityphaseSP} for details).

\begin{figure}
{\includegraphics[width=1.0\linewidth]{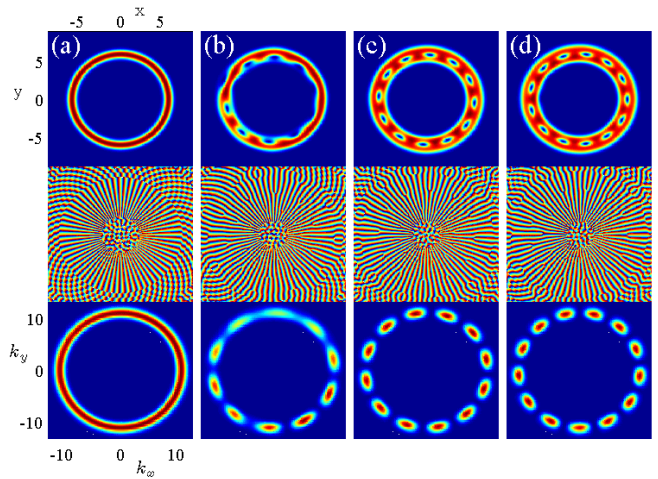}}
\caption {The profiles of the condensate wavefunctions of
the spin-up component for $\alpha=11$, $\beta=6$, and $\theta=\frac{\pi}{2}$.
The interaction parameters are $g=15$ ($a$), 35 ($b$), 75 ($c$),
and 100 ($d$), respectively. We note that (c) and (d) exhibit similar profiles but with different $q$.
From top to bottom: the density and phase profiles in real space,
and the momentum distributions which mainly are located around
the circle $|k|=\alpha$.
\label{profilepiover2}}
\end{figure}

\section{Phase transitions induced by interaction}
\label{sect:interaction}

In this section, we consider the interaction effect which will couple
single-particle eigenstates with different values of $j_z$.
It is interesting to consider the possible vortex configurations
in various parameter regimes, which has been widely considered
in the case of the fast rotating BECs.

If the dimensionless interaction parameter $g=g_{2D}N/(\hbar\omega l_T^2)$
is small, it is expected that the ground state still remains in
a giant-vortex state, which is similar to the non-interacting case.
The envelope of the variational wave-function is approximated as
\begin{eqnarray}
\phi_{j_z}(\rho,\varphi) \sim \frac{1}{2\pi^{\frac{3}{4}}\sqrt{\sigma\rho}}
e^{-\frac{(\rho-\beta)^2}{2\sigma^2}}e^{i\rho\alpha \cos\theta }
\left [ \begin{array}{c} e^{i [m\varphi-\frac{\theta}{2}]}
\\ -e^{i [(m+1)\varphi+\frac{\theta}{2}]}  \end{array} \right ] \nn
\end{eqnarray}
with $\sigma$ the radial width of the condensates.
Around a thin ring inside the cloud with the radius $\rho$, in order
to maintain the overall phase factor $e^{i m \varphi}$, the magnitude of the
local momentum along the azimuth direction is determined by
$k_{\varphi}=m/\rho$.
Depending on the width $\sigma$ of the cloud, the linewidth of
$k_{\varphi}$ is proportional to $\delta k_{\varphi}=m\sigma/\beta^2$.
In momentum space, this leads to the expansion of the distribution
around the ring with $|k|=\alpha$.
The increasing of the kinetic energy mainly comes from the term $\hat{E}_{\varphi}=(j_z/\rho-\alpha\sin\theta)^2/2$, which is estimated as
$\langle \hat{E}_{\varphi} \rangle_{j_z}$. Details derivation of various energy contributions can be found in Appendix B.

\begin{figure}
{\includegraphics[width=0.98\linewidth]{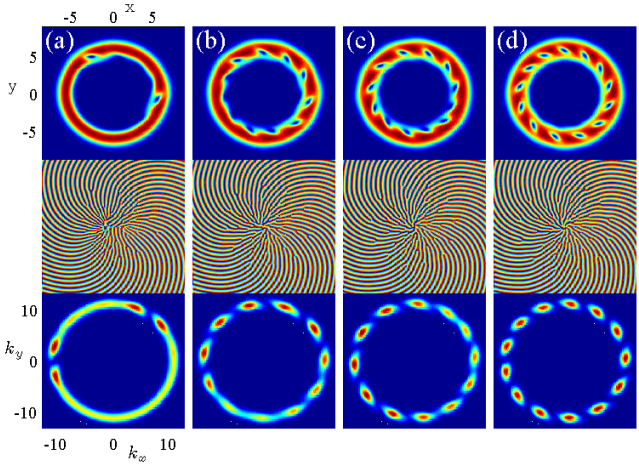}}
\caption {Ground state profiles of the condensates for $\alpha=11$,
$\beta=6$, $\theta=\pi/3$ with different interaction $g=85$(a),
$105$(b), $125$(c), and $145$(d) respectively.
From top to bottom: density and phase profiles of the spin-up component,
momentum distribution in the lower band along the circle $|k|=\alpha$.
The orientation of the ellipse-shaped vortices is determined by $\theta$.
See text for details. }
\label{profilepiover3}
\end{figure}

Increasing the interaction strength $g$ expands the cloud and leads to larger width $\sigma$ and $\delta k_{\varphi}$, which makes
the above variational state energetically unfavorable.
In order to minimize the total energy, the condensates tend to
involve additional vortices such that the local momentum mainly
distributes around the circle $|k|=\alpha$ with smaller $\delta k_{\varphi}$ .
Fig. (\ref{profilepiover2}) and (\ref{profilepiover3}) show the typical ground-state configurations
for selected parameters. 
The phase accumulations around the inwards and outwards boundaries of
the cloud are $2\pi m_+$ and $2\pi m_-$ respectively.
Therefore, there are $q=m_+-m_-$ vortices involved and distributed
symmetrically inside the condensates.
Between two nearest vortices, the local wavefunction can be approximately
determined as a plane-wave state.
Therefore, their corresponding distribution in momentum space is also
composed of $q$ peaks located symmetrically around the circle $|k|=\alpha$.

As further increase of the interaction strength, the condensates break into more pieces by involving more vortices.
The number of the vortices is qualitatively determined by the competition of the azimuthal kinetic energy and the kinetic energy introduced by the vortices.
Specifically, if $q$ vortices locate in  the middle of the cloud around the circle $\rho_0 \simeq \beta$, then for the inwards part of the condensates with $\rho < \rho_0$, the mean value of the angular momentum can be approximated as $j_{z,-}\approx j_z-q/2$, while for the regime with $\rho > \rho_0$, we have $j_{z,+}\approx j_z+q/2$.
The corresponding kinetic energy along the azimuthal direction is modified as
\begin{eqnarray}
\langle \hat{E}_{\varphi} \rangle&=&\langle \hat{E}_{\varphi} \rangle_{j_z} + \frac{q}{2\beta^2}(\frac{q}{4}-\frac{\sigma \alpha \sin\theta}{\sqrt{\pi}}).
\label{eq:energy1}
\end{eqnarray}
This indicates that, to make the vortex-lattice state favorable, we must have $(\frac{q}{4}-\frac{\sigma \alpha \sin\theta}{\sqrt{\pi}})<0$.
In the limit case with $\theta=0$, this condition is always violated.
Therefore, the ground state remains to be an eigenstate of $j_z$ with $j_z=\pm \frac{1}{2}$ even for large interaction strength.

\begin{figure}
\includegraphics[width=0.78\linewidth]{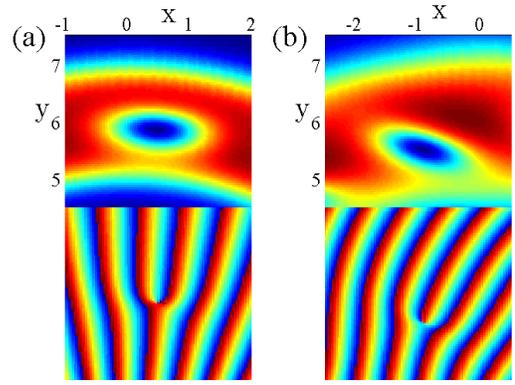}
\caption { Enlarged density and phase profiles around single vortex.
The two figures (a) and (b) are the corresponding parts adapted
from Fig. (3c) and (4d) respectively.}
\label{singlevortex}
\end{figure}

We note the vortices display an ellipse-like shape with two main axis, as shown in Fig. \ref{singlevortex}.
The phase profile is twisted, and the constant phase front exhibits a dislocation around vortex cores.
Along the direction of local wavevector $\vec{k}$, the vortex density profile is determined by the length scale $2\pi l_T/\alpha$.
While perpendicular to the direction of local $\vec{k}$, the vortex profile is dominated by the healing length $\xi$ due to interaction. 
Therefore, the vortex density distribution is determined by two different length scales in mutually orthogonal directions, which results in ellipse-shaped vortices.
Changing the interaction strength and SO coupling alerts the ratio of the two length scales, thus changes the eccentricity of the ellipses.  
Additionally, changing the angle $\theta$ also changes the direction of local magnetic fields, and thus modifies the orientation of the vortices, as shown in Fig. \ref{profilepiover2} and Fig. \ref{profilepiover3}.

On the other hand, the introduction of vortices lead to the increase of kinetic energy due to the modification of the density profile.
This can be estimated as $0.19q/(\beta\sigma \alpha \xi)$, where $\xi=1/\sqrt{2 g n_0}$ is the dimensionless healing length with $n_0$ the bulk density of the clouds.
The total energy changing due to the presence of the vortices can be written as
\begin{eqnarray}
\Delta E=\frac{q}{2\beta^2}(\frac{q}{4}-\frac{\sigma \alpha \sin\theta}{\sqrt{\pi}}) + \frac{0.19q}{\beta\sigma \alpha \xi}.
\label{eq:energy2}
\end{eqnarray}

\begin{figure}
{\includegraphics[width=0.8\columnwidth]{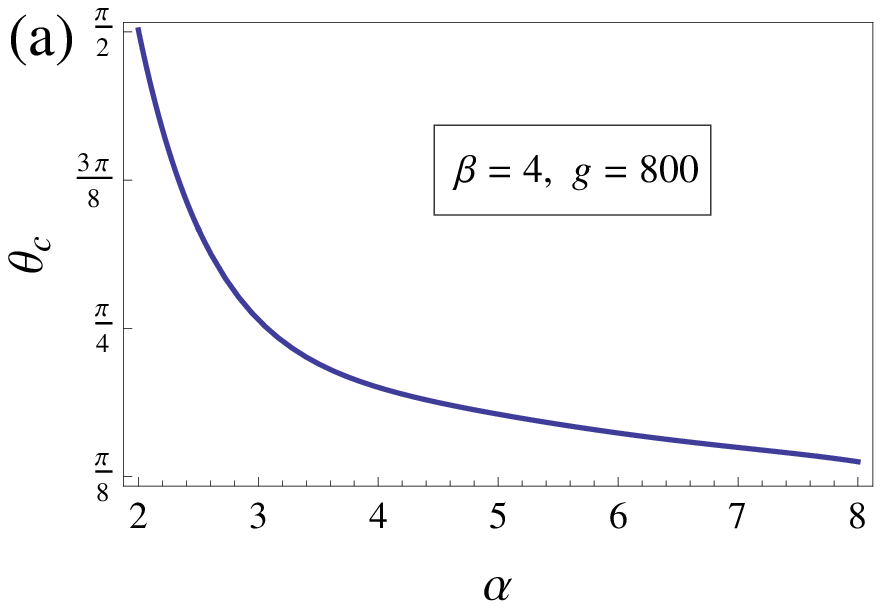}
 \includegraphics[width=0.8\columnwidth]{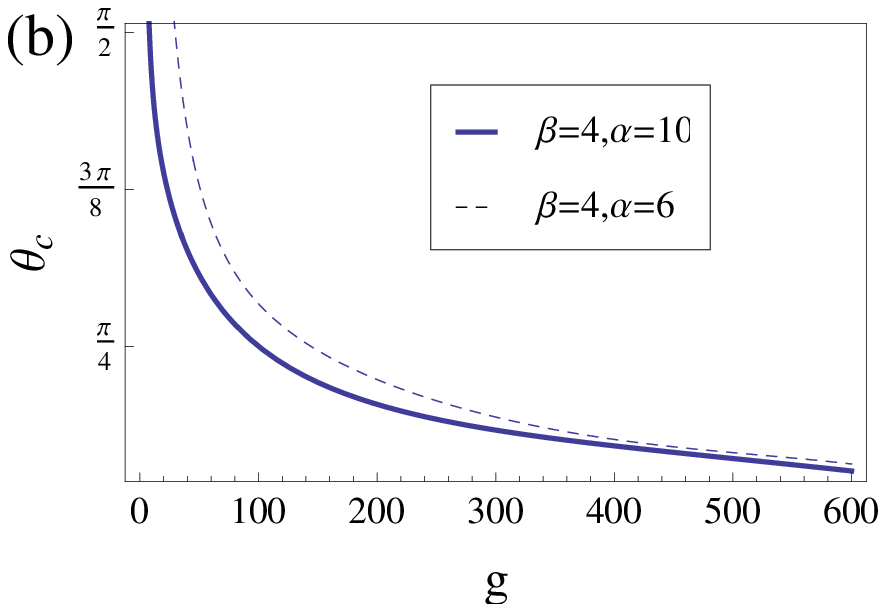}}
\caption { (a) Critical angle $\theta_c$ as a function of SO
coupling strength $\alpha$ for fixed values of $\beta=4$ and $g=800$.
(b) $\theta_c$ decreases as the increase of interaction parameter $g$
for fixed $\alpha=10$, $6$, and $\beta=4$.
\label{criticaltheta}
}
\end{figure}

Several interesting features can be extracted from Eq. \ref{eq:energy2}.
For fixed parameters $g$, $\alpha$, and $\beta$, there always exists a
critical $\theta_c$ such that $\Delta E =0$ is satisfied.
When $\theta<\theta_c$, then $\Delta E>0$, which indicates that a
giant-vortex ground state is always favored.
As increasing $\alpha$,  $\theta_c$ satisfying $\Delta E =0$ becomes smaller.
At $\theta>\theta_c$, the ground state exhibits a lattice-type
structure along the ring with a giant vortex core.
The values of $q$ is determined by minimizing $\Delta E$ with
respect to $g$, $\alpha$, and $\beta$, respectively.
Fig. (\ref{criticaltheta}) shows $\theta_c$ as a function of SO
strength $\alpha$ at which the transition from a giant-vortex state
to a vortex-lattice state occurs. 
When $\alpha$ is small, a giant-vortex state is favored for all
values of $\theta$.
As $\alpha$ increases, $\theta_c$ drops quickly initially and
decreases much slower when $\alpha$ becomes large as shown in
Fig. (\ref{criticaltheta}) (a).
In Fig. (\ref{criticaltheta}) (b), it shows that as increasing
the interaction $g$, it is becomes easier to drive the system
into the vortex lattice state.

\begin{figure}
{\includegraphics[width=0.8\columnwidth]{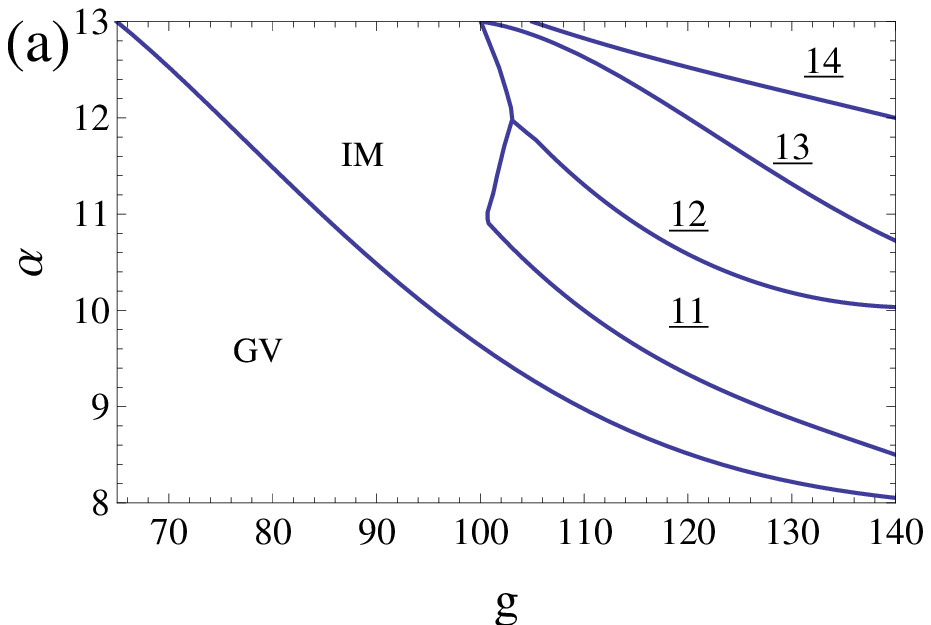}
\includegraphics[width=0.8\columnwidth]{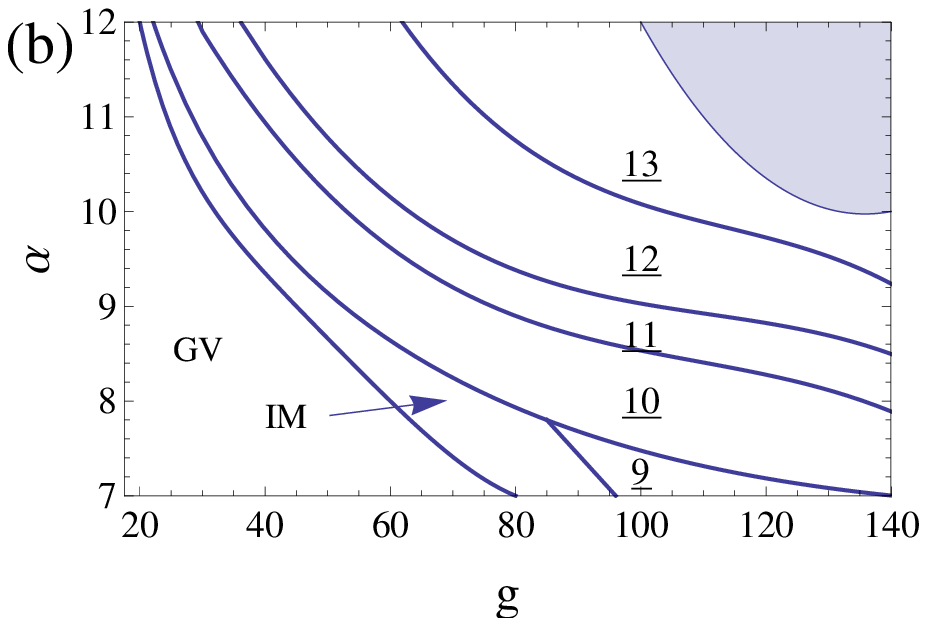}}
\caption {Phase diagram in the $\alpha-g$ plane for $\beta=6$
with different $\theta=\pi/3$ ((a)) and $\pi/2$ ((b)).
The number $\underline{M}$ means that the condensates support
a vortex-lattice-type ground state with $M$ momentum peaks along
the circle $|k|=\alpha$.
The regime with shadow in (b) indicates the ground state shows
multi-layer structure as increasing the interaction strength.
Other phases are defined as follows:
GV(giant-vortex state),IM (intermediate regime).
\label{phasediagram}
}
\end{figure}

Fig. (\ref{phasediagram}) shows the phase diagram in the $\alpha-g$
plane for a fixed $\beta=6$ for different values of $\theta$.
For a fixed $\alpha$ and at small values of $g$, the system remains to
be a giant-vortex state until $g$ reaches its critical value $g_c$.
When $g>g_c$, the system enters into an intermediate regime in which
vortices start to enter into the condensates from boundaries.
The momentum distribution also breaks into several disconnected
segments.
More single quantum vortices are generated in the condensates as further
increasing the interaction strength.
The vortices distribute symmetrically along the ring and separate
the condensates into pieces.
Between two neighboring vortices, the condensates are approximated by
local plane-wave states.
The momentum distribution composes of multi-peaks symmetrically located
around the circle $|k|=\alpha$.
Increasing $g$ also increases the number of the single quantum $M$ inside
the condensates, hence increases the number of peaks in momentum space.
For a smaller value $\theta=\pi/3$, 
the critical $g_c$ is increased, which means that stronger interactions
are needed to drive the system into the vortex-lattice states.
Interestingly, the intermediate regime is also greatly enlarged.
This is consistent with the limit case $\theta=0$, where the system
remains to be a giant-vortex state even in the case of large interaction
strength.

\begin{figure}
{\includegraphics[width=0.98\columnwidth]{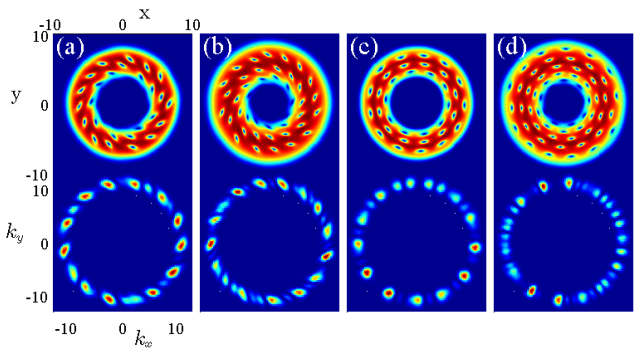}}
\caption {Density and momentum distributions about the ground states
of the condensates for $\theta=\pi/3$ with $g=400$((a)), $1000$((b)),
and $\theta=\pi/2$ with $g=400$((c)), $1000$((d)). Other parameters
are the same with figure (\ref{profilepiover3}).
We note that since the two spin components share almost the same profiles, 
only the densities of the spin-up component are shown
for simplicity. \label{multiplelayer}
}
\end{figure}

More ellipse-shaped vortices are formed as further increasing the
interaction strength, which are self-organized into a multiple
layered ring structure, as shown in Fig.(\ref{multiplelayer}).
Around each ring, vortices distributed symmetrically.
The number of the vortices between different layers can be not
equal due to their different radius.
Therefore the distribution in momentum space becomes asymmetric,
and exhibits complex multi-peak structures around the circle $|k|=\alpha$.

\section{Experimental consideration}
\label{sect:experiment}

The Hamiltonian Eq. \ref{eq:ham_1} considered above can be dynamically
generated on behalf of a series of gradient magnetic pulses
\cite{anderson2013,xu2013}.
Starting with the typical single-particle Hamiltonian
$H_s=\frac{p^2}{2m}+\frac{1}{2}m\omega^2 r^2$,
in the first time step, we employ a pair of magnetic pulses
$U_1$ and $U_1^\dagger$, defined as
as
\bea
U_1=e^{i \lambda(x \sigma_x+y \sigma_y)/\hbar},
\eea
at time $t=2n\tau$, $(2n+1)\tau$ respectively.
Secondly, a typical effective gradient coupling,
\bea
\Lambda[(x\cos\theta-y\sin\theta)\sigma_x+(y\cos\theta+x\sin\theta)\sigma_y],
\eea
is applied during the whole time duration $[(2n+1)\tau,  2(n+1)\tau]$.
Combining these two time steps, an effective dynamical evolution
$U=e^{-i H_0 \tau}$, which implements the desired dynamics.
In practice, the gradient magnetic pulse in the first cycle can
be simulated with quadrupole fields as $\vec{B}=(x,y,-2z)$.
When the condensates is strongly confined in the $xy$ plane, the
influence of the nonzero gradient along $z$-axis can be neglected.
The effective gradient coupling in the second cycle can be implemented
with the help of atom-laser coupling.
For instance, a standard two-set Raman beams with blue-detuning
\cite{liu2014} can realize an effective coupling
\bea
\Omega [\sin(\vec{k}_1\cdot \vec{r})\sigma_x+
\sin(\vec{k}_2\cdot \vec{r})\sigma_y],
\eea
where the wavevectors $\vec{k}_1$ and $\vec{k}_2$ in the $xy$ plane can
be chosen as $\vec{k}_1=k(\cos\theta, -\sin\theta)$ and $
\vec{k}_2=k(\sin\theta, \cos\theta)$.
When $2\pi/k$ is much larger than the trap length $l_T$, the required
effective coupling is approximately obtained.
Finally, the phases discussed in the context can be detected by
monitoring their corresponding density and momentum distributions
using the setup of time of flight.

\section{Concluding remarks}
\label{sect:conclusions}
To summarize, we have discussed the ground state phase diagram of
SO coupled BECs subject to gradient magnetic fields.
Theoretical and numerical analyses indicate that the system supports
various interesting vortex physics, including the single-particle
giant-vortex states with tunable vorticity, multiple
layered vortex-lattice-ring states, and the ellipse-shaped vortex profiles.
Therefore, the combination of SO coupling and the gradient magnetic
fields provides a powerful method to engineer various vortex
states without rotating the trap. We hope our work will stimulate further 
research of searching for various novel states in SO coupled bosons subject to 
effective gradient magnetic fields.

\section{Acknowledgement}
X.F. Z., Z.W. Z., and G.C. G. acknowledge the support by NSFC (Grant Nos. 11004186,
11474266,11174270), National Basic Research Program of China (Grants No. 2011CB921204 and
No. 2011CBA00200).
C. W. is supported by the NSF DMR-1410375 and AFOSR
FA9550-14-1-0168, and also acknowledges the support from the
National Natural Science Foundation of China (11328403).


\widetext
\appendix

\section{Single particle eigenstates for large $\beta$}
\label{sect:dispersion}

We start with the dimensionless Hamiltonian
\begin{eqnarray}
\frac{H_0}{\hbar\omega}=\int d \vec{\rho}^2  \hat{\phi}(\vec{\rho})^{\dag} \Big \{ -\frac{\vec{\nabla}^2}{2}  + \beta \rho \left ( \cos\theta \hat{r} +\sin\theta \hat{\varphi} \right ) \cdot \vec{\sigma} + \alpha \vec{k} \cdot \vec{\sigma} + \frac{1}{2}\rho^2 \Big \}\hat{\phi}(\vec{\rho}).
\end{eqnarray}
Since the total angular momentum is conserved, the single particle eigenstates can be written as $\phi_m=[f(\rho) e^{i m \varphi} ,
g(\rho) e^{i (m+1) \varphi}]^T$ with $j_z=(m+1/2)$. Substitute this wavefunction into their corresponding Schr\"{o}dinger equations, we obtain
\begin{eqnarray}
\left\{\frac{\hat{p}^2_{\rho}}{2} + \frac{j_z^2}{2\rho^2}  +\frac{(\rho\sigma_x-\beta)^2}{2} -\alpha [(\hat{p}_{\rho} \cos\theta + \frac{j_z}{\rho}\sin\theta ) \sigma_x \right. \nn \\
&& \hspace{-8.5cm} \left.  + (\frac{j_z}{\rho}\cos\theta-\hat{p}_{\rho} \sin\theta ) \sigma_y ] + \frac{j_z\sigma_z}{2 \rho^2}  \right \}  \left[ \begin{array}{c} \tilde{f}(\rho) \\ \tilde{g}(\rho)  \end{array} \right ]= E \left[ \begin{array}{c} \tilde{f}(\rho) \\ \tilde{g}(\rho)  \end{array} \right ], \nn
\end{eqnarray}
where $\tilde{f}(\rho)=f(\rho) e^{i\theta/2}$, $\tilde{g}(\rho)=g(\rho)e^{-i\theta/2}$, $\hat{p}_{\rho}=-i(\frac{\partial}{\partial \rho}+\frac{1}{2\rho})$ is the momentum operator along the radical direction. For large $\beta \gg 1$, these functions mainly distribute around the circle $\rho=\beta$ in the plane, so we consider the superposition $F^{\pm}(\rho)=\frac{1}{2}[\tilde{f}(\rho)\pm \tilde{g}(\rho)])$, which satisfies the following approximated equations as
\begin{eqnarray}
\left(\frac{\hat{p}^2_{\rho}}{2} \mp \alpha \cos\theta \hat{p}_{\rho}  + \frac{j_z^2}{2\rho^2} \mp \alpha \sin\theta \frac{j_z}{\rho} +\frac{\rho^2}{2} \mp \beta \rho\right ) F^{\pm}(\rho) \nn \\
\pm i \alpha \left( \hat{p}_{\rho} \sin\theta - \frac{j_z}{\rho} \cos\theta \right) F^{\mp}(\rho) = E_{j_z} F^{\pm}(\rho).  \nn
\end{eqnarray}
The above equation indicates that to minimize the kinetic energy, we need $\langle \vec{p}_{\rho} \rangle \simeq \alpha\cos\theta$.
Around $\rho=\beta$, we have the approximated solutions as $F^{\pm}(\rho) \sim H_n(\rho\pm\beta)e^{-(\rho\pm\beta)^2/2}e^{\pm i\alpha \cos\theta \rho}$ with $H_n(r)$ the usual $n$-th Hermite polynomial. Therefore, $F^+$ is negligible since we always have $\rho>0$. The solution now can be written as $\tilde{f}(\rho)\simeq\tilde{g}(\rho)\propto H_n(\rho-\beta)e^{-(\rho-\beta)^2/2}e^{i\alpha \cos\theta \rho}$. So we obtain the approximated wavefunctions for the lowest band ($n=1$) as
\begin{eqnarray}
\phi_{n=1,j_z} \simeq \frac{1}{(2\pi)^{\frac{3}{4}}\rho^{\frac{1}{2}}}e^{-\frac{(\rho-\beta)^2}{2}}e^{i\rho\alpha \cos\theta }\left [ \begin{array}{c} e^{i [m\varphi-\frac{\theta}{2}]} \\ -e^{i [(m+1)\varphi+\frac{\theta}{2}]}  \end{array} \right ].
\end{eqnarray}
The dispersion is estimated as
\begin{eqnarray}
E_{n,j_z}= n+\frac{1-\alpha^2-\beta^2}{2} + \frac{(j_z-\alpha\beta\sin\theta)^2}{2(\alpha^2\cos^2\theta+\beta^2)},
\end{eqnarray}
which is minimized when $j_z \simeq \alpha\beta\sin\theta$, so for the kinetic term along the tangential direction $\hat{E}_{\varphi}=(\frac{j_z}{\rho}-\alpha\sin\theta)^2/2$.

\section{Energy estimation of vortix lattice states around the ring}
For weak interaction, the condensates expands along the radial direction as the parameter $g$ is increased.
When $g$ is large enough, to lower the kinetic energy, the system tends to involve vortices located around a ring inside the condensates, which separate the wavefunction into two parts.
Inside the vortex-ring, the wavefunction for the spin-up component is approximated as a giant vortex with the phase factor $e^{i 2\pi m_-\phi}$, while outside the ring, the mean angular momentum carried by single particle is approximated as $m_+\hbar$.
The difference $q=m_+-m_-$ represents the vortex number inside the condensates.
Therefore the variational ground-state can be approximated as follows
\begin{eqnarray}
\phi(\rho,\phi)= \left \{
\begin{array}{c}
\phi_{j_z-q/2}(\rho,\phi) \mbox{  when  }  \rho \in (0,\beta), \\
\phi_{j_z+q/2}(\rho,\phi) \mbox{  when  }  \rho \in (\beta,\infty).
\end{array}
\right.
\end{eqnarray}
We also assumes that around the circle $\rho=\beta$, vortices are involved and self-organized to compensate the phase mismatch so that the the whole wavefunction is well-defined.

The increase of the kinetic energy mainly comes from the term $\hat{E}_{\varphi}=(\frac{j_z}{\rho}-\alpha\sin\theta)^2/2$.
For ground state without lattices, the energy is estimated and denoted as
\begin{eqnarray}
\langle \hat{E}_{\varphi} \rangle_{j_z} = \int^{\infty}_0 d\rho \int^{2\pi}_0 d\varphi \rho \phi^{\dag}_{j_z} \frac{(j_z/\rho-\alpha\sin\theta)^2}{2} \phi_{j_z},
\end{eqnarray}
where we use $\langle \rangle_{j_z}$ to denote the mean values over the trivial variational function $\phi_{j_z}$
Similarly, for GV states and large $\beta \gg 1$, the corresponding energy increasing is estimated as
\begin{eqnarray}
\langle \hat{E}_{\varphi} \rangle &=& \int^{\infty}_{\beta} d\rho \int^{2\pi}_0 d\varphi \rho \phi^{\dag}_{j_z+\frac{q}{2}} \frac{[(j_z+\frac{q}{2})/\rho-\alpha\sin\theta]^2}{2} \phi_{j_z+\frac{q}{2}} + \int^{\beta}_0 d\rho \int^{2\pi}_0 d\varphi \rho \phi^{\dag}_{j_z-\frac{q}{2}} \frac{[(j_z-\frac{q}{2})/\rho-\alpha\sin\theta]^2}{2} \phi_{j_z-\frac{q}{2}} \nn \\
&\approx& \langle \hat{E}_{\varphi} \rangle_{j_z} +\langle \frac{q^2}{8\rho^2} \rangle_{j_z} +  \int^{\infty}_{0} d\rho \int^{2\pi}_0 d\varphi \rho \phi^{\dag}_{j_z} \frac{q}{2\rho}(\frac{j_z}{\rho}-\alpha\sin\theta)  \phi_{j_z}  \nn \\
&\approx & \langle \hat{E}_{\varphi} \rangle_{j_z} + \frac{q}{2\beta^2}(\frac{q}{4}-\frac{\sigma \alpha \sin\theta}{\sqrt{\pi}}),
\end{eqnarray}
which gives the Eq. (\ref{eq:energy1}) shown in the context.

Around each vortex core, both the density and phase are twisted as shown in Fig. \ref{singlevortex}.
To take into account the energy contribution of these vortices, we choose the following approximated local wavefunction as (see Fig.(8a) for $\theta=\pi/2$)
\[
\psi \simeq \phi_0 \left[\frac{e^{-ik_+x}}{e^{-\sqrt{2}y/\xi}+1}+\frac{e^{-i(k_-x+\pi)}}{e^{\sqrt{2}y/\xi}+1}\right] \left [ \begin{array}{c} 1 \\ 1 \end{array} \right ],
\]
where for simplicity we have set the origin to the vortex core, $\phi_0$ is the bulk wavefunction away from the vortex cores and is estimated as $|\phi_0|=\sqrt{n_0}=[2\pi^{\frac{3}{4}}\sqrt{\sigma\rho}]^{-1}$, $k_{\pm}$ represent the corresponding local wavevectors of the condensates inwards and outwards the vortex ring.
$\xi=(2gn_0)^{-1/2}$ is usually called as the healing length as can be seen by considering the above wavefunction along the line $x=0$
\[
\frac{1}{e^{-\sqrt{2}y/\xi}+1} - \frac{1}{e^{\sqrt{2}y/\xi}+1} = \tanh[y/(\sqrt{2}\xi)],
\]
which is consistent with the estimation shown in \cite{book}. We note that the approximated wavefuntion is only valid when $k_-x \in [-\pi,\pi]$ and $y\in[-2\xi,2\xi]$. We also requires $k_+ \simeq 2 k_-\simeq\alpha$ such that a vortex is formed around the origin as indicated by numerics. Since energy gain along the tangential direction has been taking into account in $\hat{E}_{\varphi}$,
in this typical case, the changing of kinetic energy along the direction perpendicular to the local wavevector $\vec{k}$ can then be approximated by
\[
\int_{k_-x=-\pi}^{k_-x=\pi} dx \int_{-2\xi}^{2\xi} dy \psi^{\dag} \Big[-\frac{\nabla_y^2}{2} \Big]\psi = \frac{\sqrt{2}\tanh(\sqrt{2})^3}{3\sqrt{\pi}\beta\sigma\alpha\xi} \approx \frac{0.19}{\beta\sigma \alpha \xi}.
\]
We note that the above analysis also applies for $\theta \neq 0$. Taking into account all vortices inside the two-components condensates, we obtain the total energy introduced due to fluctuation of the vortex density profile as $0.19q/(\beta\sigma \alpha \xi)$.



\begin{thebibliography}{99} 




\bibitem{zutic2004}
I. \v{Z}uti\'{c}, J. Fabian, and S. Das Sarma, Rev. Mod. Phys. {\bf 76},  323  (2004).

\bibitem{Hasan2010}
M.Z. Hasan and C.L. Kane, Rev. Mod. Phys. {\bf 82},  3045  (2010).

\bibitem{Qi2011}
X.-L. Qi and S.-C. Zhang, Rev. Mod. Phys. {\bf 83},  1057  (2011).

\bibitem{feynman1972}
R.P. Feynman, Statistical Mechanics, A Set of Lectures (Addison-Wesley Publishing Company, ADDRESS, 1972).

\bibitem{wu2009}
C. Wu, Mod. Phys. Lett. B \textbf{23}, 1 (2009).

\bibitem{wu2011}
C. Wu, I. Mondragon-Shem, arXiv:0809.3532;
C. Wu, I. Mondragon-Shem, and X.-F. Zhou, Chin. Phys. Lett. \textbf{28}, 097102 (2011).

\bibitem{stanescu2008}
T.D. Stanescu, B. Anderson, and V. Galitski, Phys. Rev. A \textbf{78}, 023616 (2008).

\bibitem{wang2010}
C. Wang, C. Gao, C.M. Jian, and H. Zhai, Phys. Rev. Lett. \textbf{105}, 160403 (2010).

\bibitem{ho2011}
T.L. Ho and S. Zhang, Phys. Rev. Lett. \textbf{107}, 150403 (2011).

\bibitem{hu2012}
H. Hu, B. Ramachandhran, H. Pu, and X.J. Liu, Phys. Rev. Lett. \textbf{108}, 10402 (2012).

\bibitem{sinha2011}
S. Sinha, R. Nath, and L. Santos, Phys. Rev. Lett. \textbf{107}, 270401 (2011).

\bibitem{li2012}
Y. Li, X. Zhou, and C. Wu, arXiv:1205.2162 (2012).

\bibitem{kawakami2012}
T. Kawakami, T. Mizushima, M. Nitta, and K. Machida, Phys. Rev. Lett. \textbf{109}, 015301 (2012).

\bibitem{gopalakrishnan2013}
S. Gopalakrishnan, I. Martin, and E.A. Demler, Phys. Rev. Lett. \textbf{111}, 185304 (2013).


\bibitem{lin2009}
Y. Lin,  R.L. Compton, K. Jim\'{e}nez-Garc\'{\i}a, J.V. Porto1, and I.B. Spielman, Nature \textbf{462}, 628 (2009);
Y.-J. Lin, R.L. Compton, A.R. Perry, W.D. Phillips, J.V. Porto, and I.B. Spielman, Phys. Rev. Lett. \textbf{102}, 130401 (2009);
Y. Lin, K. Jimenez-Garcia, and I. Spielman, Nature \textbf{471}, 83 (2011).

\bibitem{zhang2012v1}
J.-Y. Zhang, S.-C. Ji, Z. Chen, L. Zhang, Z.-D. Du, B. Yan, G.-S. Pan, B. Zhao, Y.-J. Deng, H. Zhai, S. Chen, and J.-W. Pan, Phys. Rev. Lett. \textbf{109}, 115301 (2012).

\bibitem{wang2012}
P. Wang, Z.-Q. Yu, Z. Fu, J. Miao, L. Huang, S. Chai, H. Zhai, and J. Zhang, Phys. Rev. Lett. \textbf{109}, 095301 (2012).

\bibitem{cheuk2012}
L.W. Cheuk, A.T. Sommer, Z. Hadzibabic, T. Yefsah, W.S. Bakr, and M.W. Zwierlein, Phys. Rev. Lett. \textbf{109}, 095302 (2012)

\bibitem{qu2013}
C. Qu, C. Hamner, M. Gong, C. Zhang, and P. Engels, Phys. Rev. A \textbf{88}, 021604(R) (2013).

\bibitem{olson2014}
A.J. Olson, S.-J. Wang, R.J. Niffenegger, C.-H. Li, C.H. Greene, Y.P. Chen, Phys. Rev. A \textbf{90}, 013616 (2014).


\bibitem{wilson2013}
R.M. Wilson, B.M. Anderson, and C.W. Clark, Phys. Rev. Lett. \textbf{111}, 185303 (2013)

\bibitem{achilleos2013}
V. Achilleos, D.J. Frantzeskakis, P.G. Kevrekidis, and D.E. Pelinovsky, Phys. Rev. Lett. \textbf{110}, 264101 (2013).

\bibitem{kartashov2013}
Y.V. Kartashov, V.V. Konotop, and F.K. Abdullaev, Phys. Rev. Lett. \textbf{111}, 060402 (2013).

\bibitem{ozawa2013}
T. Ozawa and G. Baym, Phys. Rev. Lett. \textbf{110}, 085304 (2013).

\bibitem{deng2012}
Y. Deng, J. Cheng, H. Jing, C.-P. Sun, and S. Yi, Phys. Rev. Lett. \textbf{108}, 125301 (2012).

\bibitem{zhang2012}
Y. Zhang, L. Mao, and C. Zhang, Phys. Rev. Lett. \textbf{108}, 035302 (2012).

\bibitem{lobanov2014}
V. E. Lobanov, Y.V. Kartashov, and V.V. Konotop, Phys. Rev. Lett. \textbf{112}, 180403 (2014).

\bibitem{luo2014}
X. Luo, L. Wu, J. Chen, R. Lu, R. Wang, and L. You, arxiv:1403.0767v2;
L. He, A. Ji, and W. Hofstetter, arxiv:1404.0970v1;
Zeng-Qiang Yu, 1407.0990v1;
W. Han, G. Juzeli\"{u}nas, W. Zhang, and W.-M. Liu, arxiv:1407.2972v1.

\bibitem{zhou2013}
X. Zhou, Y. Li, Z. Cai, and C. Wu, J. Phys. B: At. Mol. Opt. Phys. \textbf{46}, 134001 (2013).
\bibitem{dalibard2011}
J. Dalibard, F. Gerbier, G. Juzeli\"{u}nas, and P. Ohberg, Rev. Mod. Phys. \textbf{83}, 1523 (2011).
\bibitem{zhai2014}
H. Zhai, Int. J. Mod. Phys. B. \textbf{26}, 1230001 (2012); H. Zhai, arXiv:1403.8021.
\bibitem{galistki2013}
V. Galitski and I. B. Spielman, Nature \textbf{494}, 49 (2013).
\bibitem{goldman2013}
N. Goldman, G. Juzeli\"{u}nas, P. Ohberg, I. B. Spielman, arXiv: 1308.6533.


\bibitem{anderson2013}
B.M. Anderson, I.B. Spielman, and G. Juzeli\"{u}nas, Phys. Rev. Lett. \textbf{111}, 125301 (2013).

\bibitem{xu2013}
Z.-F. Xu, L. You, and M. Ueda, Phys. Rev. A \textbf{87}, 063634 (2013).

\bibitem{kennedy2013}
C. J. Kennedy, Georgios A. Siviloglou, H. Miyake, W.C. Burton, and W. Ketterle, Phys. Rev. Lett. \textbf{111} 225301 (2013)

\bibitem{aidelsburger2013}
M. Aidelsburger, M. Atala, M. Lohse, J.T. Barreiro, B. Paredes, and I. Bloch, Phys. Rev. Lett. \textbf{111}, 185301 (2013).

\bibitem{pietila2009}
V. Pietila, and M. Mottonen,  Phys. Rev. Lett. \textbf{103}, 030401 (2009).

\bibitem{kawaguchi2008}
Y. Kawaguchi, M. Nitta, and M. Ueda, Phys. Rev. Lett. \textbf{100}, 180403 (2008).

\bibitem{schweikhard2004}
V. Schweikhard, I. Coddington, P. Engels, S. Tung, and E.A. Cornell, Phys. Rev. Lett. \textbf{93}, 210403 (2004).

\bibitem{mueller2002}
E.J. Mueller and T.-L. Ho, Phys. Rev. Lett. \textbf{88}, 180403 (2002).

\bibitem{fetter2009}
A.L. Fetter, Rev. Mod. Phys. \textbf{81}, 647 (2009).

\bibitem{zhou2011}
X. F. Zhou, J. Zhou, and C. Wu, Phys. Rev. A \textbf{84}, 063624 (2011).
\bibitem{xu2011}
X.-Q. Xu and J. H. Han, Phys. Rev. Lett. \textbf{107}, 200401 (2011).

\bibitem{radic2011}
J. Radi\'{c},  T.A. Sedrakyan, I.B. Spielman, and V. Galitski, Phys. Rev. A \textbf{84}, 063604 (2011).

\bibitem{aftalion2013}
A. Aftalion and P. Mason, Phys. Rev. A \textbf{88}, 023610 (2013).

\bibitem{fetter2014}
A.L. Fetter, Phys. Rev. A \textbf{89}, 023629 (2014).

\bibitem{liu2014}
X.-J. Liu, K.T. Law, and T.K. Ng, Phys. Rev. Lett. \textbf{112}, 086401 (2014).

\bibitem{book}  C.J. Pethick and H. Smith, Bose¨CEinstein Condensation in Dilute Gases, Chapter 7, Cambridge University Press, Cambridge, 2001.




\end{thebibliography}
\end{document}